\documentclass[aps, prd, twocolumn, preprintnumbers, letterpaper, nofootinbib, superscriptaddress, balancelastpage, 10pt]{revtex4-1}

\pdfoutput=1

\usepackage[utf8]{inputenc}

\usepackage{amsmath, amsfonts, amssymb}
\usepackage{bm, bbm}
\usepackage{enumitem}

\usepackage{hyperref}

\newcommand{\beq}{\begin{equation}\begin{aligned}{}}
\newcommand{\eeq}{\end{aligned}\end{equation}}
\newcommand{\beqa}[1]{\begin{equation}\begin{aligned}{#1}}
\newcommand{\eeqa}{\end{aligned}\end{equation}}

\newcommand{\eql}[1]{\label{eq:#1}}
\newcommand{\eq}[1]{(\ref{eq:#1})}

\newcommand{\fr}[2]{\dfrac{#1}{#2}}

\newcommand{\del}{\partial}
\newcommand{\dd}{\mathrm{d}}

\newcommand{\dt}{\!\cdot\!}

\newcommand{\too}{\longrightarrow}

\newcommand{\dsp}[1]{\displaystyle{#1}} 

\newcommand{\al}{\alpha}
\newcommand{\be}{\beta}

\newcommand{\Ga}{\Gamma}
\newcommand{\de}{\delta}
\newcommand{\De}{\Delta}
\newcommand{\ep}{\epsilon}

\newcommand{\la}{\lambda}

\newcommand{\om}{\omega}
\newcommand{\sgm}{\sigma}

\newcommand{\cL}{\mathcal{L}}
\newcommand{\cO}{\mathcal{O}}

\newcommand{\PP}{{\phantom\ast}}

\newcommand{\Lie}{\pounds}

\begin{document}

\title{Topics in gravity SCET: the diff Wilson lines and reparametrization invariance}

\author{Sabyasachi Chakraborty}
\email{sabya@hep.fsu.edu}
\affiliation{Department of Physics, Florida State University, Tallahassee, FL 32306, USA}

\author{Takemichi Okui}
\email{tokui@fsu.edu}
\affiliation{Department of Physics, Florida State University, Tallahassee, FL 32306, USA}
\affiliation{Theory Center, High Energy Accelerator Research Organization (KEK), Tsukuba 305-0801, Japan}

\author{Arash Yunesi}
\email{yunesiar@msu.edu}
\affiliation{Department of Physics, Florida State University, Tallahassee, FL 32306, USA}
\affiliation{Department of Statics and Probability, Michigan State University, East Lansing, MI 48824, USA}

\preprint{KEK-TH-2159}

\begin{abstract}
Two topics in soft collinear effective theory (SCET) for gravitational interactions are explored. First, the collinear Wilson lines---necessary building blocks for maintaining multiple copies of diffeomorphism invariance in gravity SCET---are extended to all orders in the SCET expansion parameter $\lambda$, where it has only been known to $O(\lambda)$ in the literature. Second, implications of reparametrization invariance (RPI) for the structure of gravity SCET lagrangians are studied. The utility of RPI is illustrated by an explicit example in which $O(\lambda^2)$ hard interactions of a collinear graviton are completely predicted by RPI from its $O(\lambda)$ hard interactions. It is also pointed out that the multiple diffeomorphism invariances and RPI together require certain relations among $O(\lambda)$ terms, thereby reducing the number of $O(\lambda)$ terms that need to be fixed by matching onto the full theory in the first place. 
\end{abstract}

\maketitle


\section{Introduction}
In this short note, we extend the discussions of soft collinear effective theory (SCET) for gravity presented in~\cite{Okui:2017all}.
Gravity SCET is a gravity analog of the SCET originally formulated for QCD~\cite{Bauer:2000ew, Bauer:2000yr, Bauer:2001ct, Bauer:2001yt, Bauer:2002nz}.
While the formalism of~\cite{Okui:2017all} includes both soft and collinear gravitons, 
in this note we ignore soft gravitons and exclusively focus on collinear gravitons.
Then, the relevant part of a gravity SCET lagrangian describing $N$ \emph{collinear sectors}%
\footnote{i.e., $N$ well-collimated energetic ``jets'' widely separated in angle from one another, 
where a ``jet'' may refer to particles in the initial state.
Fields in the $n^\text{th}$ collinear sector consist only of Fourier modes pointing approximately in the direction of the $n^\text{th}$ ``jet axis'' within a small angle of $O(\la)$.}
can be written in the following form:
\beq
\cL_\text{SCET} = \sum_{n=1}^N \cL_n + \cL_\text{hard}
\,,
\eeq
where $\cL_n$ consists only of fields in the $n^\text{th}$ collinear sector,
while all \emph{hard interactions}---the interactions that couple different collinear sectors---are in $\cL_\text{hard}$.
Using $\la$ as a measure of collinearness (i.e., $\la \sim p_\perp / E \ll 1$),
$\cL_\text{SCET}$ is an expansion in powers of $\la$.
$\cL_n$ can be straightforwardly obtained by expanding the full-theory lagrangian to the desired order in $\la$ 
(e.g., see~\cite{Beneke:2012xa} for some first worked out examples of $\cL_n$ at $O(\la)$).
All the meat of the effective theory is in $\cL_\text{hard}$.    
Referring to the dimension of $\cL_\text{hard}$ in the absence of gravity as $O(\la^0)$, 
Ref.~\cite{Okui:2017all} worked out, for the first time, how to construct $\cL_\text{hard}$ at $O(\la)$.
It also showed that the absence of collinear singularities in gravity---first investigated by Weinberg~\cite{Weinberg:1965nx} and later proven rigorously in the full theory by Akhoury, Saotome and Sterman~\cite{Akhoury:2011kq}---follows trivially from an effective gauge symmetry of gravity SCET 
in that there are simply no operators at $O(\la^n)$ with $n\leq 0$ that respect the symmetry.

The first problem we will discuss in this note concerns this effective gauge symmetry. 
As in QCD SCET~\cite{Bauer:2001ct, Bauer:2001yt}, the splitting of a gauge field into $N$ collinear sectors means that we effectively have $N$ copies of the corresponding gauge invariance, one for each collinear sector.
For gravity SCET, this means that we have $N$ copies of diffeomorphism (diff) invariance and local Lorentz symmetry, 
where transformation parameters for the $n^\text{th}$ diff$\times$Lorentz group are restricted to only have the Fourier modes of the $n^\text{th}$ collinear sector.
Then, writing an operator in $\cL_\text{hard}$ as $\cO_1 \cO_2 \cdots \cO_N$ with $\cO_n$ consisting only of fields in the $n^\text{th}$ collinear sector, 
each $\cO_n$ must be gauge invariant under the $n^\text{th}$ collinear diff$\times$Lorentz because all other $\cO_m$ with $m \neq n$ are trivially invariant under the $n^\text{th}$ collinear diff$\times$Lorentz.
Analogously to QCD SCET, Ref.~\cite{Okui:2017all} accomplishes such invariance by introducing Wilson lines, i.e., \emph{the collinear diff Wilson line} for collinear diff invariances and \emph{the collinear Lorentz Wilson line} for collinear local Lorentz invariance.
(These collinear Wilson lines should not be confused with the ``collinear Wilson line'' studied in~\cite{Beneke:2012xa} as the latter is not a Wilson line for either diff or Lorentz invariance but was introduced as a re-summation of certain $O(1/\la)$ couplings.)
While \cite{Okui:2017all} gives an exact closed-form expression of the collinear Lorentz Wilson line, 
it provides an expression of the collinear diff Wilson line only to $O(\la)$. 
In this note, we will present a method that permits us to express the collinear diff Wilson line to \emph{any desired order in $\la$.}

The second problem we investigate is \emph{reparametrization invariance} (RPI) in gravity SCET\@. 
As in QCD SCET, the $n^\text{th}$ collinear sector is most conveniently described in terms of \emph{the $n^\text{th}$ lightcone coordinates,} which is defined such that the components of a collinear momentum $q$ in this sector should scale in $\la$ as
\beq
(q^{+_n}, q^{-_n}, q^{i_n}) 
= (q_{-_n}^\PP, q_{+_n}^\PP, -q_{i_n}^\PP)
\sim (\la^0, \la^2, \la)
\eql{collinear_momentum}
\eeq
with $i_n = 1_n, 2_n$ referring to the two spatial directions orthogonal to the ``jet axis'' of the $n^\text{th}$ collinear sector. 
The metric in this coordinate system is indicated above in the relation between the upper and lower indices, 
which in particular implies ${\dsp q_1 \dt q_2} \sim \la^2$ if both $q_1$ and $q_2$ obey the scaling law~\eq{collinear_momentum}.
On the other hand,
the assumption that different collinear sectors are widely separated from one another means that
\beq
(p^{+_n}, p^{-_n}, p^{i_n}) 
= (p_{-_n}^\PP, p_{+_n}^\PP, -p_{i_n}^\PP)
\sim (\la^0, \la^0, \la^0)
\eql{hard_momentum}
\eeq
for any $p$ that belongs to an $m^\text{th}$ collinear sector with $m \neq n$.
So, we have $\dsp{p_1 \dt p_2 \sim \la^0}$ if $p_1$ and $p_2$ belong to different collinear sectors. 
Now, \emph{reparametrization} (RP) refers to a change in the bases of lightcone coordinates that keeps both scaling laws~\eq{collinear_momentum} and~\eq{hard_momentum} unchanged.
Since any two EFTs with identical field contents, symmetries, and power counting laws describe the same physics, 
our theory must be RP invariant (RPI)~\cite{Chay:2002vy, Manohar:2002fd}. 
As we will discuss below, despite being a change in coordinate bases, 
RP differs from diff$\times$ Lorentz in terms of scaling laws. 
Thus, RPI can indeed lead to additional constraints on the structure of the lagrangian beyond diff$\times$Lorentz.
We will illustrate the utility of RPI in gravity SCET by deriving all $O(\la^2)$ operators from $O(\la)$ operators in $\cL_\text{hard}$ for the scattering of two scalars into two scalars plus a collinear graviton at tree level.

\section{Collinear diff Wilson lines to all orders}
We begin with the collinear diff Wilson lines.
Since a collinear Wilson line only concerns a single collinear sector, 
we drop the index ${}_n$ on the lightcone coordinates
and simply refer to them by $+$, $-$, and $i=1,2$.
Throughout this note,
we adopt the convention that we convert all diff indices to Lorentz indices by using (inverse) vierbeins so that all operators are diff scalars.
Ref.~\cite{Okui:2017all} then gives the collinear diff Wilson line, $V(x)$, to $O(\la)$:
\beq
V_1(x) = 1 - \Bigl\{ \fr{1}{\del_-^2} \Ga^\mu_{--}(x) \Bigr\} \del_\mu
\,,\eql{V:O(lambda)} 
\eeq
where the ``${}_1$'' on $V_1$ indicates that it is an $O(\la^0) + O(\la)$ expression, while
\beq
\fr{1}{\del_-} f(x^-) \equiv \int_{-\infty}^{x^-} \!\! \dd x^{-\prime} \, f(x^{-\prime})
\,,
\eeq
and $\{\cdots\}$ indicates that the integration only acts on the expression between $\{$ and $\}$.%
\footnote{This Wilson line evidently comes from an infinite past and is appropriate when it acts on an operator $\phi(x)$ that \emph{annihilates} collinear particles.
When $\phi(x)$ \emph{creates} collinear particles, $1 / \del_-$ should instead be interpreted as
$\dsp{-\int_{x^-}^\infty \!\! \dd x^{-\prime} \, f(x^{-\prime})}$ so that the Wilson line would extend to an infinite future.
Rephrasing the following discussion for this case is a trivial matter.}
When $V_1$ acts on an arbitrary diff scalar operator $\phi(x)$ of collinear modes, 
the product $\dsp{V_1\phi}$ is diff invariant to $O(\la)$, i.e., 
$\de(V_1\phi) = O(\la^2)$ under $x^\mu \to x^\mu - \xi^\mu(x)$ with an infinitesimal $\xi^\mu$.
It is useful to re-verify this to lay the groundwork for establishing $V$ to all orders.
We need to know two things for this purpose.
First, under diff, we have $\de\phi = \xi^\mu \del_\mu\phi$ and
\beq
\de\Ga^\mu_{\nu\rho} = \del_{\nu} \del_{\rho} \xi^\mu + \Lie_\xi \Ga^\mu_{\nu\rho}
\,,\eql{delta_Gamma}
\eeq
where $\Lie_\xi$ denotes the Lie derivative with respect to $\xi^\mu$:
\beq
\Lie_\xi \Ga^\mu_{\nu\rho}
&= \xi^\sgm \del_\sgm \Ga^\mu_{\nu\rho}
\\ &\phantom{=\;} 
- (\del_\sgm \xi^\mu) \Ga^\sgm_{\nu\rho} + (\del_\nu \xi^\sgm) \Ga^\mu_{\sgm\rho} + (\del_\rho \xi^\sgm) \Ga^\mu_{\nu\sgm}
\,.\eql{Lie}
\eeq
Second, we need some power counting rules from~\cite{Okui:2017all}:
\beq
\del_\mu \sim q_\mu
\,,\quad
\xi^\mu \sim \fr{q^\mu}{\la}
\,,\quad
\Ga^\mu_{\nu\rho} \sim \fr{q^\mu q_\nu q_\rho}{\la}
\,,\eql{scaling:collinear_diff}
\eeq
where $q$ is a collinear momentum, scaling as in~\eq{collinear_momentum}.
These relations tell us that the Christoffel term in~\eq{V:O(lambda)} is indeed $O(\la)$
as $(1 / q_-^2) (q^\mu q_- q_- / \la) q_\mu \sim \la$.
For the variation of $\Ga^\mu_{\nu\rho}$ in~\eq{delta_Gamma}, 
the first term on the right-hand side is the same order in $\la$ as $\Ga^\mu_{\nu\rho}$ itself
while the Lie derivative term is higher order by one power of $\la$,
which can be seen from~\eq{scaling:collinear_diff} as $q_\nu q_\rho q^\mu / \la \sim \Ga^\mu_{\nu\rho}$
and $(q^\sgm / \la) q_\sgm \Ga^\mu_{\nu\rho} \sim \la \Ga^\mu_{\nu\rho}$, respectively.
The scaling laws~\eq{scaling:collinear_diff} also tell us that $\de\phi = \xi^\mu \del_\mu\phi \sim \la\phi$.
Then, in $\de(V_1\phi)$, we only have two $O(\la)$ contributions: 
$\de\phi$, and the term containing $\de\Ga^\mu_{--}$ with the Lie derivative term ignored.
These two $O(\la)$ contributions cancel with each other, so $\dsp{V_1\phi}$ is indeed collinear diff invariant to $O(\la)$.

We now describe a systematic way to find $V$ to any desired order in $\la$.
First, we define the collinear diff Wilson line $V$ acting on a collinear diff scalar $\phi$ as
\beq
V\!\phi(x) \equiv \phi(X)
\,,\eql{def:V}
\eeq
where $X^\mu$ is the coordinates of the geometrical point that would be equal to $x^\mu$ in the absence of gravity.
Being a geometrical point, the \emph{actual geometrical location} of point $X$ in spacetime is diff invariant even though the \emph{coordinates} representing its location change under diff.
Therefore, the value of $\phi$ at this location, $\phi(X)$, is diff invariant.

To draw such $X$ in spacetime,  
we identify $X$ as the endpoint of a semi-infinite geodesic $\bar{x}^\mu(s)$:
\beq
X^\mu \equiv \bar{x}^{\mu} (s) \bigr|_{s=0}
\,.\eql{def:X}
\eeq
In the absence of gravity, this endpoint would be at $x^\mu$ and the geodesic would be directed in the $x^-$ direction%
\footnote{As usual, the reason for picking out the $x^-$ direction is due to the power counting~\eq{collinear_momentum},
which says powers of $\del_-$ are unsuppressed and thereby allows local operators to be displaced from one another in the $x^-$ direction.}
toward an infinite past ($s \to -\infty$).
Therefore, we take the geodesic equation
\beq
\fr{\dd^2 \bar{x}^\mu}{\dd s^2} = -\Ga^\mu_{\nu\rho}(\bar{x}) \, \fr{\dd \bar{x}^\nu}{\dd s} \fr{\dd \bar{x}^\rho}{\dd s}
\,,\eql{geo_eqn}
\eeq
and solve for $\bar{x}^\mu(s)$ perturbatively in powers of $\Ga^\mu_{\nu\rho}$ as
$\bar{x}^\mu(s) = \bar{x}_0^\mu(s) + \bar{x}_1^\mu(s) + \bar{x}_2^\mu(s) + \cdots$
($\bar{x}_n^\mu \sim O(\Ga^n)$)
with the ``initial'' condition:
\beq
\bar{x}_0^\mu \Bigr|_{s=0} = x^\mu
\,,\quad
\fr{\dd \bar{x}_0^\mu}{\dd s} \biggr|_{s=0} = \de^\mu_-
\,.
\eeq
The integration constants for $x^\mu_n$ with $n \geq 1$ should be chosen such that $x^\mu(s) \to x_0^\mu(s)$ as $s \to -\infty$, reflecting the boundary condition that no gravity was present in the infinite past. 
Needless to say, we also expand $X^\mu$ as $X^\mu = X_0^\mu + X_1^\mu + X_2^\mu + \cdots$ ($X_n^\mu \sim O(\Ga^n)$).

To illustrate how it works explicitly, let us find $V$ to the second order and directly verify that $\de(V_2\phi) = O(\la^3)$.
First, at $O(\Ga^0)$, we have
\beq
\bar{x}_0^\mu = x^\mu + \de^\mu_{-} s
\,,\quad
X_0^\mu = x^\mu
\,,
\eeq
as we should.
Plugging this back into~\eq{geo_eqn} and picking up $O(\Ga^1)$ terms, we get
\beq
\fr{\dd^2 \bar{x}^\mu_1}{\dd s^2} = -\Ga^\mu_{--}(\bar{x}_0)
\quad\Longrightarrow\quad
\bar{x}_1^\mu = -\Bigl\{ \fr{1}{\del_s^2} \Ga^\mu_{--}(\bar{x}_0) \Bigr\}
\,.
\eeq
Evaluating $\bar{x}_1^\mu$ at $s=0$ then gives
\beq
X_1^\mu = -\Bigl\{ \fr{1}{\del_-^2} \Ga^\mu_{--}(x) \Bigr\}
\quad\sim \fr{q^\mu}{\la}
\,,
\eeq
where~\eq{scaling:collinear_diff} was used to get the scaling.
Substituting $X_0 + X_1$ for $X$ in~\eq{def:V} and expanding it to $O(\la)$ indeed reproduces $V_1$ in~\eq{V:O(lambda)}.

Moving on to $O(\Ga^2)$ terms, we have
\beq
\fr{\dd^2 \bar{x}^\mu_2}{\dd s^2} 
&= -2\Ga^\mu_{\nu-} \, \fr{\dd \bar{x}_1^\nu}{\dd s} 
-\bigl( \del_\nu \Ga^\mu_{--} \bigr) \, \bar{x}_1^\nu
\\
&=
2\Ga^\mu_{\nu-} \Bigl\{ \fr{1}{\del_s} \Ga^\nu_{--} \Bigr\}
+ \bigl( \del_{\nu} \Ga^\mu_{--} \bigr) \Bigl\{ \fr{1}{\del_s^2} \Ga^\nu_{--} \Bigr\}
\,,
\eeq
where all the Christoffel symbols are evaluated at point $\bar{x}_0^\mu$ and all $\bar{x}_{0,1,2}^\mu$ at an arbitrary $s$.  
This leads to
\beq
X_2^\mu 
&= \biggl\{\! \fr{1}{\del_-^2} \!\biggl[ 
2\Ga^\mu_{\nu-} \Bigl\{ \fr{1}{\del_-} \Ga^\nu_{--} \Bigr\}
+ \bigl( \del_{\nu} \Ga^\mu_{--} \bigr) \Bigl\{ \fr{1}{\del_-^2} \Ga^\nu_{--} \Bigr\} \!
\biggr] \! \biggr\}
\,,
\eeq
where all the Christoffel symbols are now evaluated at point $x$.
From~\eq{scaling:collinear_diff}, we see that
\beq
X_2^\mu \sim q^\mu
\,,
\eeq
so $X_2$ is higher order than $X_1$ by one power of $\la$ as it should be.

We can now substitute $X_0 + X_1 + X_2$ for $X$ in~\eq{def:V} to obtain $V_2$ and verify that $\de(V_2\phi) = O(\la^3)$.
First, expanding in $\la$ to second order, we have
\beq
V_2\phi(x) 
&= \phi(x) + X_1^\mu \del_\mu \phi(x) 
\\ &\phantom{=\;}
+ \Bigl( X_2^\mu 
+ \fr12 X_1^\mu X_1^\nu \del_\nu \Bigr) \del_\mu \phi(x)
\,.\eql{V:O(lambda^2)}
\eeq
Then, referring to~\eq{scaling:collinear_diff} and ignoring terms of $O(\la^3)$ and higher, we get
\beq
\de(V_2\phi) 
&= \de\phi + \de X_1^\mu \, \del_\mu \phi + X_1^\mu \, \del_\mu \de\phi
\\ &\phantom{=\;}
+ ( \de X_2^\mu + X_1^\mu \, \de X_1^\nu \, \del_\nu ) \, \del_\mu \phi
\\ 
&=  -\Bigl\{ \fr{1}{\del_-^2} \Lie_{\xi} \Ga^\mu_{--} \Bigr\} \del_\mu \phi + X_1^\mu \, \del_\mu \de\phi
\\ &\phantom{=\;}
+ ( \de X_2^\mu + X_1^\mu \, \de X_1^\nu \, \del_\nu ) \, \del_\mu \phi 
\,,\eql{deltaVphi_calculation}
\eeq
with $\de\phi = \xi^\rho \del_\rho\phi$. 
In the last expression above, all terms are $O(\la^2)$ as all $O(\la)$ terms had cancelled out. 
For the $\de X_1^\mu$ and $\de X_2^\mu$ in the last line of~\eq{deltaVphi_calculation},
we should simply use $\de \Ga^\mu_{\nu\rho} = \del_\nu \del_\rho \xi^\mu$, 
because the Lie derivative term would result in $O(\la^3)$ terms.
It is then straightforward to verify that all the terms in~\eq{deltaVphi_calculation} cancel out, 
demonstrating that $\de(V_2\phi) = O(\la^3)$ indeed.

The process above can clearly be iterated to an arbitrary order in $\la$.
We therefore have a method to compute collinear diff Wilson lines to any desired order in $\la$.

\section{Reparametrization invariance}
We now move on to our second topic, i.e., RPI\@.
Let us write a general RP transformation as $p_\mu \to p'_\mu = p_\mu + p_\nu \,\om^\nu_{~\mu}$ with $\om^\nu_{~\mu}$ being infinitesimal.
Demanding that the lightcone metric as indicated in~\eq{collinear_momentum} be invariant under RP, we get $\omega_\mu^{~\nu} = -\om^\nu_{~\mu}$.
Despite this property of $\om^\nu_{~\mu}$, RP is not Lorentz transformations nor 
a subset of diff transformations, because RP transformation parameters must obey different scaling properties from the Lorentz/diff transformation parameters. 
As derived in~\cite{Okui:2017all}, if $\om^\nu_{~\mu}$ were a Lorentz or diff transformation parameter, 
it would scale as $\omega^\nu_{~\mu} \sim q^\nu q_\mu / \la$, where $q$ is a collinear momentum scaling as in~\eq{collinear_momentum}. 
Then, for a momentum $p$ scaling as in~\eq{hard_momentum},  
we would have $p'_\mu \>\sim\> p_\mu + p_\nu \, (q^\nu q_\mu / \la) \>\sim\> p_\mu + q_\mu / \la$.
But this would give $p'_- \sim 1 / \la$, contradicting with the original scaling, $p_-^\PP \sim \la^0$.
Therefore, RP must be given a different scaling law from diff$\times$Lorentz transformations. 
This means that RPI would constrain the SCET lagrangian differently from diff$\times$Lorentz, 
thereby offering us an additional predictive power. 

Following the RPI nomenclature in QCD SCET, 
we distinguish three types of RP\@.
Type-I RP~\cite{Chay:2002vy} is generated by a 3-vector parameter $\De^{\!i} = -\De_i \equiv \om^i_{~+}$,
which give
\beq
p'_+ = p_+^\PP + \De^{\!i} p_i^\PP
\,,\quad
p'_- = p_-^\PP 
\,,\quad
p'_i = p_i^\PP - \De_i \, p_-^\PP
\,.\eql{Type-I}
\eeq
Demanding that the scaling law be preserved when $p$ is collinear to $q$, we get $\De^{\!i} \sim \la$.
The same demand on a non-collinear $p$ would only give a weaker condition.
Type-II~\cite{Manohar:2002fd} is generated by $\ep^i = -\ep_i \equiv \om^i_{~-}$, which give
\beq
p'_+ = p_+^\PP 
\,,\quad
p'_- = p_-^\PP + \ep^i p_i^\PP
\,,\quad
p'_i = p_i^\PP - \ep_i \, p_+^\PP
\,.\eql{Type-II}
\eeq
Demanding that the scaling law be preserved when $p$ is not collinear to $q$, we get $\ep^i \sim \la^0$.
The same demand on a collinear $p$ would only give a weaker condition.
Type-III~\cite{Manohar:2002fd} is generated by $\al \equiv \omega^+_{~+}$, leading to $p'_+ = (1 + \al) p_+$, $p'_- = (1 - \al) p_-$, and $p'_i = p_i^\PP$ with $\al \sim \la^0$.
Type-III is literally a Lorentz boost in the lightcone direction, for which the theory is already invariant.
So, Type-III does not give additional constraints beyond diff$\times$Lorentz.
Similarly, the theory is already manifestly invariant under rotations generated by $\om^i_{~j}$. 
However, both Type-I and -II RPs can give us nontrivial constraints on the structure of a gravity SCET lagrangian that is not implied by diff$\times$Lorentz.

To be concrete, we consider a simple example of scattering of two massless scalars into two massless scalars plus a graviton, where the graviton is collinear to one of the initial scalars. 
That is, we have four collinear sectors ($N=4$), and consider $\phi(p_1) + \phi(p_2) \to \phi(p_3) + \phi(p_4) + h_{\mu\nu}(q)$ with $q \sim p_1$.
We assume that the only non-gravitational coupling in the full theory is $\cL_\text{int} = -\phi^4 / 4!$, 
and we do not consider loops.
As worked out in~\cite{Okui:2017all}, 
the $O(\la^0)$ and $O(\la)$ interactions in $\cL_\text{hard}$ are given by
\beq
\cL_\text{hard}^{(0+1)} 
&= -(V_1\phi_1^\PP) \, \phi_2^\PP \phi^*_3 \phi^*_4
\\ &\phantom{=\;}
-\fr12 \Bigl\{ \fr{1}{\del_-^2} R_{--} \Bigr\} \phi_1^\PP \phi_2^\PP \phi^*_3 \phi^*_4 
\\ &\phantom{=\;}
+\Bigl\{ \fr{1}{\del_-^3} R_{-i-j} \Bigr\} \phi_1^\PP \Bigl\{ \fr{\del^i\del^j}{\del_+} \phi_2^\PP \Bigr\} \phi^*_3 \phi^*_4 + (2 \to 3, 4) 
\,,\eql{NLP}
\eeq
where $V_1$ is the $O(\la)$ collinear diff Wilson line~\eq{V:O(lambda)}, 
while $(2 \to 3, 4)$ indicates the same operators as the preceding one with a Riemann tensor except that $\phi_2$ is replaced by $\phi^*_3$ or $\phi^*_4$. 
Here and below, spacetime indices always refer to the $1^\text{st}$ lightcone coordinates, where $\phi_1$ and the graviton by definition belong to the $1^\text{st}$ collinear sector.

In~\eq{NLP}, $V_1$ contains both $O(\la^0)$ and $O(\la)$ terms, while the Ricci and Riemann terms are both $O(\la)$ because $R_{--} \sim R_{-i-j} \sim \la$~\cite{Okui:2017all}.
The $O(\la)$ term inside $\dsp{V_1\phi_1}$ is related by symmetry to the $O(\la^0)$ term, 
while the coefficients of the Ricci and Riemann terms are determined by matching onto the full theory.
This is the first place where gravity SCET is far more practically efficient and conceptually transparent than the full theory.
Since $h_{\mu\nu}$ scales as $q_\mu q_\nu / \la$~\cite{Okui:2017all}, 
the scaling law~\eq{collinear_momentum} tells us that $h_{--} \sim 1 / \la$.
Hence, in the full theory calculation of the amplitude for the $h_{--}$ polarization, 
one must painfully expand each propagator of each diagram to $O(\la^2)$, 
only to find that all $O(1 / \la)$ and $O(\la^0)$ terms cancel out after adding up all diagrams.
In the gravity SCET calculation, in contrast, symmetry and power counting tell us without any calculations 
that there are simply no operators containing a collinear graviton at $O(1 / \la)$ and $O(\la^0)$.
Hard interactions manifestly begin at $O(\la)$ in the SCET lagrangian.
Symmetry and power counting also tell us that all $O(\la)$ terms beside the ones from collinear Wilson lines must come with $R_{--}$ or $R_{-i-j}$~\cite{Okui:2017all}. 
To fix the forms of the $R_{--}$ and $R_{-i-j}$ terms,
we can just choose to match the simplest polarizations like $h_{ij}$ or $h_{-+}$, 
as they are already $O(\la)$ and we do not need any further $\la$ expansions on the full theory side.

Moving now onto $O(\la^2)$, we find that RPI gives strong constraints on the structure of $O(\la^2)$ operators and determine them completely.
First, notice that extending diff$\times$Lorentz to $O(\la^2)$ does not predict any $O(\la^2)$ interactions relevant for our process in question, $\phi_1 + \phi_2 \to \phi_3 + \phi_4 + h_{\mu\nu}$  at tree level.
This is because the process only involves one graviton, 
while the $O(\la^2)$ terms arising from multiplying the terms in~\eq{NLP} by vierbeins or Wilson lines for diff$\times$Lorentz would be all 2-graviton terms.
We thus disregard those $O(\la^2)$ terms.   
For the same reason, we also ignore the $O(\la^2)$ terms corresponding to 2-graviton terms from $R_{\mu\nu}$, $R_{\mu\nu\rho\sigma}$, $(R_{\mu\nu})^2$, etc., 
as well as the $O(\la^2)$ terms in $V_2$ we derived earlier as they are all quadratic in $h_{\mu\nu}$. 
Therefore, for the purpose of seeing what $O(\la^2)$ interactions are required by the symmetries of the theory for $\phi_1 + \phi_2 \to \phi_3 + \phi_4 + h_{\mu\nu}$ at tree level, 
we only need to promote~\eq{NLP} to make it RPI\@.

Let us begin with RPI-promoting the expression~\eq{V:O(lambda)}.
It is already invariant under Type-I as $\del_-$ does not transform under Type-I (see~\eq{Type-I}).
It is also already invariant under Type-III as the transformations of the two lower-minus indices in the numerator are cancelled by the two lower-minus indices in the denominator.
It is not RPI under Type-II, however,
as the lower-minus index does transform under Type-II (see~\eq{Type-II}).

To promote the expression~\eq{V:O(lambda)} to an RPI form, we employ and extend a strategy proposed in~\cite{Marcantonini:2008qn} for QCD SCET.
The basic idea is the following.
Imagine a collinear momentum $q$ and another momentum $p$ in a different collinear sector from $q$.
Then, at the leading order in $\la$, we have $\dsp{q \dt p} = q_- p_+ \sim \la^0$.
Here, the product $q_- p_+$ is not RPI beyond $O(\la^0)$ but the original full dot product, $\dsp{q \dt p}$, is RPI to all orders as it does not refer to any specific choice of lightcone coordinate basis.
Thus, we can RPI-promote $q_- p_+$ by changing it back to $\dsp{q \dt p}$ and re-expanding it to the desired order in $\la$. 
While this technique is used for RPI-promoting Wilson lines in~\cite{Marcantonini:2008qn},
we will be applying it to the hard interactions as well.

To use this strategy for RPI-promoting~\eq{V:O(lambda)}, we introduce a notation that allows us to handle different collinear sectors simultaneously.
In the new notation, \eq{V:O(lambda)} is rewritten as 
\beq
V_1 = 
1 - \fr{1}{({}_0\del_-)^2} \, \Ga^\mu_{--} \, {}_1\del_\mu
\,,\eql{V:O(lambda):new}
\eeq
where the left subscript on a derivative indicates the field it acts on.
Namely, ${}_0\del$ acts only on $h_{\mu\nu}$,  
while ${}_1\del$, \ldots, ${}_4\del$ act only on $\phi_1$, \ldots, $\phi_4$, respectively.
On the other hand, regardless of what left subscripts refer to,
right super/subscripts always refer to the $1^\text{st}$ lightcone coordinates as we mentioned already.
Thus, for example, the power counting laws~\eq{collinear_momentum} and~\eq{hard_momentum} tell us that ${}_{0,1}\del_+ \sim \la^2$ and ${}_{2,3,4}\del_+ \sim \la^0$.

With this notation, we can now promote~\eq{V:O(lambda):new} to an RPI form using the strategy describe above:
\beq
V_1 \too\> 
V_1^\text{RPI} 
= 
1 - \fr13 \sum_{n=2}^4
\fr{{}_n\del^\nu \, {}_n\del^\rho}{({}_n\del \cdot {}_0\del)^2} \, \Ga^\mu_{\nu\rho} \, {}_1\del_\mu
\,,\eql{Wilson:RPI}
\eeq
To see how $V_1^\text{RPI}$ reduces to~\eq{V:O(lambda):new} at $O(\la)$, 
let us consider the $n=2$ term in~\eq{Wilson:RPI} for example. 
Using the power counting laws~\eq{collinear_momentum}, \eq{hard_momentum}, and~\eq{scaling:collinear_diff},
we get
\beq
\fr{{}_2\del^\nu \, {}_2\del^\rho}{( {}_2\del \cdot {}_0\del )^2} \, \Ga^\mu_{\nu\rho} \, {}_1\del_\mu
&= \fr{{}_2\del_+ \, {}_2\del_+}{({}_2\del_+ \, {}_0\del_-)^2} \, \Ga^\mu_{--} \, {}_1\del_\mu 
+ O(\la^2)
\\ 
&= \fr{1}{( {}_0\del_- )^2} \, \Ga^\mu_{--} \, {}_1\del_\mu 
+ O(\la^2)
\,.
\eeq
Thus, together with the factor of $1/3$ and the sum over $n=2,3,4$, we indeed recover the expression~\eq{V:O(lambda):new}.
The sum over $n$ with an identical coefficient in~\eq{Wilson:RPI} is not required by RPI 
but is necessary for crossing symmetry of the amplitude.

Now that $V_1^\text{RPI}$ is manifestly RPI, expanding $V_1^\text{RPI}$ to $O(\la^2)$ will give us $O(\la^2)$ terms predicted by RPI given the presence of the $O(\la)$ collinear diff Wilson line.
There are two types of the $O(\la^2)$ contributions. 
First, 
for $(\nu,\rho) = (-,i)$ or $(i,-)$ in~\eq{Wilson:RPI}, 
the $\Ga^\mu_{\nu\rho} \, {}_1\del_\mu$ factor is already $O(\la^2)$ as we can see from the scaling law~\eq{scaling:collinear_diff}.
Second, for $(\nu,\rho) = (-,-)$, the $\Ga^\mu_{\nu\rho} \, {}_1\del_\mu$ factor is $O(\la)$ but we can also pick up an $O(\la)$ term from the Taylor expansion of the denominator:
\beq
\fr{1}{{}_n\del \cdot {}_0\del}
= \fr{1}{{}_n\del_+ \, {}_0\del_-} 
\biggl( 1 - \fr{{}_n\del^i \, {}_0\del_i}{{}_n\del_+ \, {}_0\del_-} + O(\la^2) \biggr)
\,,\eql{denominator}
\eeq
where the second term inside the parentheses is $O(\la)$.
Combining these two types of $O(\la^2)$ contributions, we find that the $O(\la^2)$ part of $V_1^\text{RPI}$ is
\beq
-\fr23 \!\sum_{n=2}^4 \biggl[ 
\fr{{}_n\del^i}{{}_n\del_+ \, ({}_0\del_-)^2} \, \Ga^\mu_{i-} 
-\fr{{}_n\del^i \, {}_0\del_i}{{}_n\del_+ \, ({}_0\del_- )^3} \, \Ga^\mu_{--} 
\biggr] {}_1\del_\mu
\,,
\eeq
which can further be combined into a single term:
\beq
\fr23 \!\sum_{n=2}^4 
\fr{{}_n\del^i \, {}_1\del^\mu}{{}_n\del_+ \, ({}_0\del_-)^3} \, 
R_{\mu-i-}
\,.\eql{Wilson:NNLP}
\eeq
This acting on $-\phi_1^\PP \phi_2^\PP \phi_3^* \phi_4^*$ thus gives the $O(\la^2)$ hard interactions predicted from RPI-promoting the Wilson line term in~\eq{NLP}.

The Ricci term in~\eq{NLP} can also be promoted to an RPI form in the same way via the replacement
\beq
\fr{1}{\del_-^2} R_{--}
\too\>
\fr13 \sum_{n=2}^4 \fr{{}_n\del^\mu \, {}_n\del^\nu}{( {}_n\del \cdot {}_0\del )^2} R_{\mu\nu}
\,.\eql{Ricci:RPI}
\eeq
To expand this in $\la$, we need to recall the power counting law for the Ricci tensor~\cite{Okui:2017all}:
\beq
R_{\mu\nu} \sim \la q_\mu q_\nu
\,.\eql{power:Ricci}
\eeq
We again have two types of $O(\la^2)$ contributions as we did from $V_1^\text{RPI}$ above.
We find that the $O(\la^2)$ part of the RPI-promoted expression in~\eq{Ricci:RPI} is
\beq
\fr23 \sum_{n=2}^4 \biggl[ 
\fr{{}_n\del^i}{{}_n\del_+ \, ({}_0\del_-)^2} \, R_{i-}
- \fr{{}_n\del^i \, {}_0\del_i}{{}_n\del_+ \, ({}_0\del_-)^3} \, R_{--} 
\biggr]
\,.\eql{Ricci:NNLP}
\eeq
This acting on $-\frac12\phi_1^\PP \phi_2^\PP \phi_3^* \phi_4^*$ gives the $O(\la^2)$ interactions predicted by RPI-promoting the Ricci term in~\eq{NLP}.

RPI also works for the Riemann term in~\eq{NLP} in essentially the same way.
With our left subscript notation, the $O(\la)$ Riemann term in~\eq{NLP} can be rewritten as
\beq
\sum_{n=2}^4 \fr{{}_n\del^i \, {}_n\del^j }{{}_n\del_+ \, ({}_0\del_-)^3} \, R_{-i-j} \, 
\phi_1^\PP \phi_2^\PP  \phi^*_3 \phi^*_4
\,.\eql{Riemann:NLP}
\eeq
This can be RPI-promoted as
\beq
\fr13 \sum_{(m,n)}
\fr{ {}_m\del^\al \, {}_n\del^\mu \, {}_m\del^\be \,  {}_n\del^\nu}{( {}_m\del \cdot {}_0\del )^2 \, ( {}_n\del \cdot {}_0\del )} \, R_{\al\mu\be\nu} \, 
\phi_1^\PP \phi_2^\PP  \phi^*_3 \phi^*_4
\,,\eql{Riemann:RPI}
\eeq
where $m,n =2,3,4$ and ``$(m,n)$'' indicates summation over all $m$ and $n$ with $m \neq n$,
with $(n,m)$ being regarded as distinct from $(m,n)$.
Let us first verify that the expression~\eq{Riemann:RPI} reduces to~\eq{Riemann:NLP} at $O(\la)$.
From the power counting law for the Riemann tensor~\cite{Okui:2017all}:
\beq
R_{\mu\nu\rho\sgm} \sim \fr{q_{[\mu} q_{\nu]} q_{[\rho} q_{\sgm]}}{\la}
\,,\eql{power:Riemann}
\eeq
we see that the $O(\la)$ terms in~\eq{Riemann:RPI} arise from picking up $O(\la)$ terms from the numerator by setting $(\al,\mu,\be,\nu)$ $=$ $(-,i,-,j)$, $(i,-,j,-)$, $(-,i,j,-)$, or $(i,-,-,j)$,
while picking up $O(\la^0)$ contributions from the denominator by truncating it to $({}_m\del_+ \, {}_0\del_-)^2 \, ({}_n\del_+ \, {}_0\del_-)$.
Let us describe how the $n=2$ term of~\eq{Riemann:NLP} arises from this.
First, adding up the $(-,i,-,j)$ terms coming from the $(m,n) = (3,2)$ and $(4,2)$ cases gives $2/3$ of the $n=2$ term of~\eq{Riemann:NLP}.
Next, adding up the $(i,-,j,-)$ terms from the $(m,n) = (2,3)$ and $(2,4)$ cases gives $-1/3$ of the $n=2$ term of~\eq{Riemann:NLP} as
\beq
\fr13 \fr{R_{i-j-}}{({}_0\del_-)^3} \, \fr{{}_2\del^i \, {}_2\del^j}{( {}_2\del_+ )^2} \, \bigl[ {}_3\del_+ + {}_4\del_+ \bigr]
=
-\fr13 \fr{R_{-i-j}}{({}_0\del_-)^3} \, \fr{{}_2\del^i \, {}_2\del^j}{( {}_2\del_+ )^2} \, {}_2\del_+
\,.\eql{-1/3}
\eeq
Here we have used the momentum conservation, ${}_0\del_\mu +  {}_1\del_\mu + \cdots + {}_4\del_\mu = 0$, where ${}_0\del_+$ and ${}_1\del_+$ are negligible as ${}_{0,1}\del_+ \sim \la^2$ and ${}_{2,3,4}\del_+ \sim \la^0$ from the scaling laws~\eq{collinear_momentum} and~\eq{hard_momentum}.
Finally, the $(-,i,j,-)$ and $(i,-,-,j)$ terms from the $(m,n) = (2,3)$ and $(2,4)$ cases give $2/3$ of the $n=2$ term of~\eq{Riemann:NLP} as
\beq
-\fr23 \fr{R_{-i-j}}{({}_0\del_-)^3} \, \fr{{}_2\del^i}{{}_2\del_+} \, \bigl[ {}_3\del^j + {}_4\del^j \bigr]
=
\fr23 \fr{R_{-i-j}}{({}_0\del_-)^3} \, \fr{{}_2\del^i}{{}_2\del_+} \, {}_2\del^j
\,,\eql{2/3}
\eeq
where momentum conservation has again been used with ${}_{0,1}\del^j \sim \la$ being neglected compared to ${}_{2,3,4}\del^j \sim \la^0$.
Adding up all these contributions gives us the $n=2$ term of~\eq{Riemann:NLP}.
We thus see that the RPI expression~\eq{Riemann:RPI} indeed reproduces the correct $O(\la)$ terms in~\eq{Riemann:NLP}.

Our next task is to expand~\eq{Riemann:RPI} to $O(\la^2)$ to get $O(\la^2)$ interactions predicted by RPI given the $O(\la)$ Riemann term in~\eq{NLP}.
We have two sources of $O(\la^2)$ contributions depending on whether we directly get $O(\la^2)$ components of $R_{\al\mu\be\nu}$ or pick up an $O(\la)$ component of $R_{\al\mu\be\nu}$ and multiplying it by an $O(\la)$ combination of derivatives.
The latter case is further divided into two categories depending on whether the $O(\la)$ combination of derivatives comes from the $O(\la)$ term in the expansion of the denominator~\eq{denominator} 
or from the ${}_{0,1}\del^j$ thrown away in~\eq{2/3}.
Let us go through these one-by-one.

First, from~\eq{power:Riemann},
the only $O(\la)$ component of $R_{\al\mu\be\nu}$ is $R_{-i-j}$ up to obvious permutations of the indices, 
while its $O(\la^2)$ components are $R_{-ijk}$ and $R_{-+-i}$ up to permutations.
So, the direct $O(\la^2)$ contributions from $R_{\al\mu\be\nu}$ are given by the following:
\begin{itemize}
\item{
$R_{-ijk}$ ($\sim O(\la^2)$) multiplied by the leading order denominators.
Adding up all relevant permutations of the indices and using momentum conservation in a similar way as we did for~\eq{2/3}, we get
\beq
\fr23 \sum_{(m,n)} 
\fr{{}_n\del^i \, {}_m\del^j \, {}_n\del^k}{{}_n\del_+ \, {}_m\del_+ \, {}_0\del_-}
\fr{R_{-ijk}}{({}_0\del_-)^2} 
\,.\eql{Riemann:NNLP2}
\eeq
}
\item{
$R_{-+-i}$ ($\sim O(\la^2)$) multiplied by the leading order denominators.
Here, again after using momentum conservation, we obtain
\beq
\fr23 \sum_{n=2}^4 \!\left[
3 \, \fr{{}_n\del_- \, {}_n\del^i}{{}_n\del_+ \, {}_0\del_-} 
+\fr{{}_n\del^i}{{}_n\del_+} 
+\fr{{}_n\del^i \, {}_1\del_-}{{}_n\del_+ \, {}_0\del_-}
\right]\! \fr{R_{-+-i}}{({}_0\del_-)^2}
\,.\eql{Riemann:NNLP3}  
\eeq
}
\end{itemize}
On the other hand, the $O(\la^2)$ contributions from an $O(\la)$ combination of derivatives acting on $R_{-i-j} \sim O(\la)$ are the following:
\begin{itemize}
\item{
$R_{-i-j}$ multiplied by the $O(\la)$ term from a denominator as in~\eq{denominator}.
After momentum conservation, 
the $(\al,\mu,\be,\nu)=(-,i,-,j)$ terms of~\eq{Riemann:RPI} become
\beq
-\fr23
\biggl[ \sum_{(m,n)}\! \fr{{}_m\del^k}{{}_m\del_+}                         
          +\sum_{n=2}^4 \fr{{}_n\del^k}{{}_n\del_+}
\biggr]
\fr{{}_n\del^i \, {}_n\del^j \, {}_0\del_k}{{}_n\del_+ \, ({}_0\del_-)^2} 
\fr{R_{-i-j}}{({}_0\del_-)^2} 
\,,\eql{Riemann:NNLP1.1}
\eeq
while the $(i,-,j,-)$ terms become
\beq
\sum_{n=2}^4 
\fr{{}_n\del^i \, {}_n\del^j \, {}_n\del^k \, {}_0\del_k}{({}_n\del_+)^2 \, ({}_0\del_-)^2}
\fr{R_{-i-j}}{({}_0\del_-)^2} 
\,,\eql{Riemann:NNLP1.2}
\eeq
and the $(i,-,-,j)$ and $(-,i,j,-)$ terms together give
\beq
\fr23
\!\biggl( \sum_{(m,n)}\! \fr{{}_m\del^i}{{}_m\del_+}                         
          -2\sum_{n=2}^4 \fr{{}_n\del^i}{{}_n\del_+}
\biggr)
\fr{{}_n\del^j \, {}_n\del^k \, {}_0\del_k}{{}_n\del_+ \, ({}_0\del_-)^2} 
\fr{R_{-i-j}}{({}_0\del_-)^2} 
\,.\eql{Riemann:NNLP1.3}
\eeq
}
\item{
The ${}_{0,1}\del^j$ terms neglected in~\eq{2/3}:
\beq
\fr23 \sum_{n=2}^4 \fr{R_{-i-j}}{({}_0\del_-)^3} \, \fr{{}_n\del^i}{{}_n\del_+} \, \bigl[ {}_0\del^j + {}_1\del^j \bigr]
\,.\eql{Riemann:NNLP4}
\eeq
}
\end{itemize}
Now, in adding up all the contributions above, 
notice that the Bianchi identity implies that $R_{-ijk}$ in~\eq{Riemann:NNLP2} can be rewritten as
\beq
R_{-ijk}
= \fr{{}_0\del_j}{{}_0\del_-} R_{-i-k} - \fr{{}_0\del_k}{{}_0\del_-} R_{-i-j}
\,.\eql{Bianchi}
\eeq
Then, the first and second terms on the right-hand side here cancel out with the first terms of~\eq{Riemann:NNLP1.1} and~\eq{Riemann:NNLP1.3}, respectively. 
No double summations, $\sum_{(m,n)}$, remain. 
Only the single summations, $\sum_n$, survive.
Moreover, the ${}_1\del_-$ term of~\eq{Riemann:NNLP3} and the ${}_1\del^j$ term of~\eq{Riemann:NNLP4} together cancel the $O(\la^2)$ contributions~\eq{Wilson:NNLP} from the Wilson line.
None of these cancellations is actually an accident as we will discuss later.
The total $O(\la^2)$ interactions predicted by RPI-promiting $\cL_\text{hard}^{(0+1)}$ are, therefore, given by
\beq
\cL_\text{hard, RPI}^{(2)} 
&= \sum_{n=2}^4 \biggl[
\fr13
\!\!\left( 
\fr{{}_n\del^i \, {}_0\del_i}{{}_n\del_+ \, {}_0\del_-} \fr{R_{--}}{({}_0\del_-)^2}
-\fr{{}_n\del^i}{{}_n\del_+ \, {}_0\del_-} \fr{R_{i-}}{{}_0\del_-}
\right)
\\
& \hspace{3ex}
+\fr23
\!\!\left(
3 \, \fr{{}_n\del_- \, {}_n\del^i}{{}_n\del_+ \, {}_0\del_-} 
+\fr{{}_n\del^i}{{}_n\del_+} 
\right)\!\! \fr{R_{-+-i}}{({}_0\del_-)^2}
\\
& \hspace{3ex}
+\!\left( 
\fr23 \fr{{}_n\del^i \, {}_0\del^j}{{}_n\del_+ \, {}_0\del_-}
-\fr{{}_n\del^i \, {}_n\del^j \, {}_n\del^k \, {}_0\del_k}{({}_n\del_+)^2 \, ({}_0\del_-)^2}
\right)\!\! \fr{R_{-i-j}}{({}_0\del_-)^2}
\biggr] 
.\eql{NNLP}
\eeq

Our next task is to show that there are no $O(\la^2)$ operators other than~\eq{NNLP}.
For this purpose, it is useful to first understand why $O(\la)$ operators must be in the form~\eq{NLP} up to numerical coefficients.
The Wilson line term, of course, is required and determined by gauge invariance with no associated free parameter.
The rest of $\cL_\text{hard}^{(0+1)}$ must be also gauge invariant.
Because of the power counting laws~\eq{power:Ricci} and~\eq{power:Riemann}, 
the only $O(\la)$ components of gauge covariant objects are $R_{--}$ and $R_{-i-j}$.
For our process, which only has one graviton emission, these components are also gauge invariant, 
as multiplying them by Wilson lines to make them gauge invariant would only produce multi-graviton couplings.
In $\cL_\text{hard}^{(0+1)}$, whatever is in front of $R_{--}$ or $R_{-i-j}$ must be dimensionless,
so each of $R_{--}$ and $R_{-i-j}$ must be divided by two derivatives.
The two derivatives must be ${}_n\del_- \, {}_m\del_-$ to ``cancel'' the two $-$ indices in $R_{--}$ or $R_{-i-j}$ for Type-III RPI\@.
To proceed further, we need to just imagine gross features of the matching calculation as follows. 
To determine $n$ and $m$ for our process of a single graviton emission at tree level, 
observe that a derivative in the denominator in $\cL_\text{hard}$ arises only when $\phi_2$, $\phi_3$, or $\phi_4$ emits the graviton in a full-theory diagram. 
For example, when $\phi_2$ emits the graviton in a full-theory diagram, the internal $\phi_2$ propagator between the graviton emission and the $\phi^4$ vertex goes as
\beq
\fr{1}{(p_2^\PP - q)^2} \propto \fr{1}{p_2^\PP \dt q}
= \fr{1}{p_{2+}^\PP q_-^\PP} \biggl( 1 + \fr{p_2^i \, q_i^\PP}{p_{2+}^\PP q_-^\PP} + \fr{p_{2-}^\PP q_+^\PP}{p_{2+}^\PP q_-^\PP} \biggr)^{\!\! -1}
\,,
\eeq
where the second and third terms inside the parentheses are $O(\la)$ and $O(\la^2)$, respectively.
We thus see that Taylor-expanding propagators only give powers of $1 / q_-$, never $1 / p_{n-}$ with $n=1, \ldots, 4$, so we must have $n=m=0$ for the ${}_n\del_- \, {}_m\del_-$.
We thus have the combinations $R_{--} / ({}_0\del_-)^2$ and $R_{-i-j} / ({}_0\del_-)^2$.
For the latter, the $i$ and $j$ indices must be contracted with $\dsp{{}_n\del^i \, {}_m\del^j}$,
and to make it dimensionless we must divide it by a Type-III invariant product of two derivatives, 
which as we just learned above must be $\dsp{{}_\ell\del_+ \, {}_0\del_-}$.
But at tree level with only one graviton, we must have $n=m=\ell$ because in each full-theory diagram the graviton couples to only one of $\phi_2$, $\phi_3$, and $\phi_4$.
And all $\phi_2$, \ldots, $\phi_4$ should appear symmetrically so we must sum over $n$ with the same coefficient.
Therefore, the only possible forms are
\beq
\fr{R_{--}}{({}_0\del_-)^2} 
\,,\quad
\sum_{n=2}^4 \fr{{}_n\del^i \, {}_n\del^j}{{}_n\del_+ \, {}_0\del_-} \fr{R_{-i-j}}{({}_0\del_-)^2} 
\,.\eql{NLP_operators}
\eeq
There are no more combinations of derivatives that could be inserted into these.
Whatever they are, they must be ratios of derivatives to be dimensionless.
As we have seen above, 
the only $O(\la^0)$ Type-III invariant combination that could appear in the denominator is $\dsp{{}_n\del_+ \, {}_0\del_-}$.
For each such denominator,
the numerator has to be dimension-2, $O(\la^0)$, and Type-III invariant, 
and the only such combinations are $\dsp{{}_n\del_+ \, {}_0\del_-}$, $\dsp{{}_n\del_+ \, {}_n\del_-}$, and $\dsp{{}_n\del^i \, {}_n\del_i}$.
The first one would just cancel the denominator. 
The second and third ones actually have to be added together with a common coefficient for the Lorentz invariance of the collinear sector $n$, thereby forming $\dsp{{}_n\del^\mu \, {}_n\del_\mu}$.
But this acting on $\phi_n$ vanishes by the leading-order equation of motion.
We thus see that the operators in~\eq{NLP_operators} are the only possible $O(\la)$ operators for $\cL_\text{hard}$ with one collinear graviton emission at tree level.

Let us follow the same line of reasoning for $O(\la^2)$.
For Ricci, it must be either $R_{--} / ({}_0\del_-)^2$ multiplied by an $O(\la)$ dimensionless ratio of derivatives, or $R_{-i} / {}_0\del_-$ multiplied by an $O(\la^0)$ ratio of dimension $-1$ carrying an index $i$.
The latter can only be $\dsp{{}_n\del^i / ({}_n\del_+ \, {}_0\del_-)}$ with $n=2,3,4$,
because $\dsp{{}_m\del^i / ({}_n\del_+ \, {}_0\del_-)}$ with $m \neq n$ cannot appear at tree level 
while $\dsp{{}_n\del^i / ({}_n\del_+ \, {}_0\del_-)}$ with $n=0,1$ would be $O(\la)$ rather than $O(\la^0)$.
Similarly, the former can only be $\dsp{{}_n\del^i \, {}_0\del_i / ({}_n\del_+ \, {}_0\del_-)}$.
For both cases, there are no more combinations of derivatives that could be inserted without vanishing by the leading-order equation of motion.
Thus, for Ricci, the only possible $O(\la^2)$ operators are
\beq
\sum_{n=2}^4 \fr{{}_n\del^i \, {}_0\del_i}{{}_n\del_+ \, {}_0\del_-} \fr{R_{--}}{({}_0\del_-)^2} 
\,,\quad
\sum_{n=2}^4 \fr{{}_n\del^i}{{}_n\del_+ \, {}_0\del_-} \fr{R_{-i}}{{}_0\del_-} 
\,.
\eeq
These are precisely the Ricci operators we have in~\eq{NNLP}, and we have already seen that their coefficients are fixed by Type-II RPI\@.

Similarly, for Riemann, a similar reasoning leads to only four possibilities:
\beq
&\sum_{n=2}^4 
\fr{{}_n\del^i \, {}_0\del^j}{{}_n\del_+ \, {}_0\del_-} \,
\fr{R_{-i-j}}{({}_0\del_-)^2} 
\,,\quad
\sum_{n=2}^4 
\fr{{}_n\del^i \, {}_n\del^j}{{}_n\del_+ \, {}_0\del_-} 
\fr{{}_n\del^k \, {}_0\del_k}{{}_n\del_+ \, {}_0\del_-} \, 
\fr{R_{-i-j}}{({}_0\del_-)^2} 
\,,\\
&\sum_{n=2}^4 
\fr{{}_n\del^i \, {}_n\del_-}{{}_n\del_+ \, {}_0\del_-} 
\fr{R_{-+-i}}{({}_0\del_-)^2} 
\,,\quad
\sum_{n=2}^4 
\fr{{}_n\del^i}{{}_n\del_+} 
\fr{R_{-+-i}}{({}_0\del_-)^2} 
\,.\eql{Riemann_list}
\eeq
These are precisely the Riemann operators we have in~\eq{NNLP}, and we have already seen that their coefficients are fixed by Type-II RPI\@.

We now see that the cancellations of $R_{-ijk}$ terms due to the Bianchi identity~\eq{Bianchi} were actually destined to occur.
Observe the absence of $R_{-ijk}$ in the list~\eq{Riemann_list}.
This follows from the fact that ${}_n\del^i {}_n\del^j \, {}_n\del^k$ acting on $R_{-ijk}$ would vanish by the antisymmetry in $j$ and $k$,
while ${}_n\del^i \, {}_0\del^j \, {}_n\del^k$ on $R_{-ijk}$ would be $O(\la^3)$.
Thus, the $R_{-ijk}$ term from~\eq{Riemann:NNLP2} had to cancel.

To conclude this section, we have seen that RPI completely fixes the $O(\la^2)$ operators given the $O(\la)$ operators~\eq{NLP} as
\beq
\cL_\text{hard}^{(2)} 
= \cL_\text{hard, RPI}^{(2)} 
\,,
\eeq
where $\cL_\text{hard, RPI}^{(2)}$ is given in~\eq{NNLP}.
These operators are very compact and can be readily translated into the amplitude at $O(\la^2)$.
Like the $O(\la)$ case, or even worse, the corresponding full-theory calculation of the $O(\la^2)$ terms is very ``inefficient'' with many ``unexpected'' cancellations, but   
we have explicitly checked that it agrees with the EFT amplitude.
We thus see that RPI is a powerful and useful tool in gravity SCET\@.

\section{Summary and Discussions}
In this note, we first derived a formula from which collinear diff Wilson lines can be computed to an arbitrary order in $\la$. 
This is a necessary ingredient in gravity SCET as the hard interaction, $\cL_\text{hard}$, in the effective lagrangian must be invariant under collinear diff gauge groups. 

Next, we discussed RPI and illustrated how RPI can significantly constrain the structure of the effective lagrangian by working out an example in which $O(\la^2)$ interactions in $\cL_\text{hard}$ are completely fixed by RPI from given $O(\la)$ interactions.  

It appears that RPI can actually do even more---it can even reduce the amount of matching calculation at $O(\la)$. This is suggested in our specific example by the cancellation of the $O(\la^2)$ terms~\eq{Wilson:NNLP} arising from RPI-promoting the collinear diff Wilson line (recall the discussion above the expression~\eq{NNLP}).
Notice that the expression~\eq{Wilson:NNLP} involves other collinear sectors than the $1^\text{st}$ collinear sector. 
But from the perspective of gauge symmetry, a collinear Wilson line should only involve fields within its own collinear sector.
Thus, we could have foreseen that the contributions~\eq{Wilson:NNLP} would be cancelled by some other contributions. 
We can turn this around and \emph{demand} that the numerical factor and derivative structures acting on the $R_{-i-j}$ at $O(\la)$ be such that the $O(\la^2)$ terms arising from it by RPI cancel those from the Wilson line. 
This completely fixes the $O(\la)$ Riemann term in~\eq{NLP} without any $O(\la)$ matching calculation onto the full theory. 
This leaves us only the Ricci term in~\eq{NLP} to be matched at $O(\la)$. 
Once that is done, all $O(\la^2)$ hard interactions can be derived from RPI without any matching calculation (except for the very gross features of full-theory diagrams we used) as we have seen above.

Therefore, when combined with other symmetries and some gross properties of amplitudes, RPI can lead to strong constraints on the structures of effective lagrangians in gravity SCET\@.         
Our discussions above, however, also tell us what complications we should expect when we go beyond tree level and/or the 1-graviton emission.
For example,
at loop level, a graviton propagator in a full theory diagram can connect different collinear sectors.
This suggests that our argument above that led to the absence of double or triple summations becomes invalid and we should expect multiple summations.
Therefore, a dedicated study akin to what was done in~\cite{Larkoski:2014bxa} for the soft theorems in QCD SCET must be also done for gravity SCET to see the power and utility of RPI beyond tree level for gravitational amplitudes.

Moreover, the cancellations of the $R_{-ijk}$ terms and $O(\la^2)$ Wilson line terms from RPI, both of which are compulsory as we have discussed above, may be an indication that our implementation of RPI is perhaps not optimal as it introduces terms that we know will cancel at the end.
Therefore, even restricting ourselves at tree level and 1-graviton emission, we have an interesting problem of figuring out such optimal implementation of RPI in gravity SCET\@.

It should be emphasized that constraints from RPI are in addition to those from diff$\times$Lorentz. 
By using the Wilson lines, an effective Lagrangian can be made diff$\times$Lorentz invariant to any desired order in $\lambda$,
although the invariance is not \emph{manifest} due to the use of the lightcone coordinates.
RPI gives additional constraints arising from removing the reference to the lightcone coordinates.  

Finally, an important topic left out in this note is the invariance under the RP between soft and collinear modes.
Once soft modes are reintroduced to the theory, shifting collinear momenta by soft momenta does not affect the power counting of collinear modes and thus constitutes a redundancy under which the theory should be invariant~\cite{Marcantonini:2008qn}. 
It is very possible that this RPI can lead to further constraints on the hard interactions at $O(\la^2)$, which is the lowest order at which both the soft and collinear interactions are present in gravity SCET~\cite{Okui:2017all}.
We leave this very interesting problem for future work.

\section*{Acknowledgment}
This work was supported by the US Department of Energy grant DE-SC0010102. 
This work was performed in part at Aspen Center for Physics supported by National Science Foundation grant PHY-1607611,
and also in part at the Kavli Institute for Theoretical Physics supported by the National Science Foundation under Grant No.~NSF PHY-1748958.


%

\end{document}